\documentclass[trackchanges]{aastex701}

\begin{document}

\title{Self-constrained Magnetic Reconnection of a Solar Flare Co-spatial with a Non-ejected Filament}

\author[0000-0002-1190-0173]{Ye Qiu}
\affiliation{Institute of Science and Technology for Deep Space Exploration, Suzhou Campus, Nanjing University, Suzhou 215163, China}
\affiliation{Key Laboratory of Modern Astronomy and Astrophysics (Nanjing University), Ministry of Education, Nanjing 210023, China}
\affiliation{Key Laboratory of Space Weather, National Satellite Meteorological Center (National Center for Space Weather), China Meteorological Administration, Beijing 100081, China}
\email{qiuye@nju.edu.cn}

\author{Changxue Chen}
\affiliation{Key Laboratory of Dark Matter and Space Astronomy, Purple Mountain Observatory, Chinese Academy of Sciences, Nanjing 210023, China}
\affiliation{School of Astronomy and Space Science, University of Science and Technology of China, Hefei 230026, China}
\email{cxchen@pmo.ac.cn}

\author{Anqin Chen}
\affiliation{Key Laboratory of Space Weather, National Satellite Meteorological Center (National Center for Space Weather), China Meteorological Administration, Beijing 100081, China}
\affiliation{Innovation Center for FengYun Meteorological Satellite (FYSIC), Beijing 100081, China}
\email{chenanqin@cma.gov.cn}

\author[0000-0001-7693-4908]{Chuan Li}
\affiliation{Institute of Science and Technology for Deep Space Exploration, Suzhou Campus, Nanjing University, Suzhou 215163, China}
\affiliation{School of Astronomy and Space Science, Nanjing University, Nanjing 210023, China }
\affiliation{Key Laboratory of Modern Astronomy and Astrophysics (Nanjing University), Ministry of Education, Nanjing 210023, China}
\email{lic@nju.edu.cn}

\author[0000-0002-9293-8439]{Yang Guo}
\affiliation{School of Astronomy and Space Science, Nanjing University, Nanjing 210023, China }
\affiliation{Key Laboratory of Modern Astronomy and Astrophysics (Nanjing University), Ministry of Education, Nanjing 210023, China}
\email{guoyang@nju.edu.cn}

\author[0000-0003-1350-9722]{Linggao Kong}
\affiliation{Institute of Science and Technology for Deep Space Exploration, Suzhou Campus, Nanjing University, Suzhou 215163, China}
\affiliation{Key Laboratory of Modern Astronomy and Astrophysics (Nanjing University), Ministry of Education, Nanjing 210023, China}
\email{lgkong@nju.edu.cn}

\author{Feng Lu}
\affiliation{Key Laboratory of Radiometric Calibration and Validation for Environmental Satellites, National Satellite Meteorological Center (National Center for Space Weather), China Meteorological Administration, Beijing 100081, China}
\affiliation{Innovation Center for FengYun Meteorological Satellite (FYSIC), Beijing 100081, China}
\email{lufeng@cma.gov.cn}

\author{Xiaohu Zhang}
\affiliation{Key Laboratory of Radiometric Calibration and Validation for Environmental Satellites, National Satellite Meteorological Center (National Center for Space Weather), China Meteorological Administration, Beijing 100081, China}
\affiliation{Innovation Center for FengYun Meteorological Satellite (FYSIC), Beijing 100081, China}
\email{zhangxiaohu@cma.gov.cn}

\author{Weiguo Zong}
\affiliation{Key Laboratory of Space Weather, National Satellite Meteorological Center (National Center for Space Weather), China Meteorological Administration, Beijing 100081, China}
\affiliation{Innovation Center for FengYun Meteorological Satellite (FYSIC), Beijing 100081, China}
\email{zongwg@cma.gov.cn}

\author{Jinsong Wang}
\affiliation{Key Laboratory of Space Weather, National Satellite Meteorological Center (National Center for Space Weather), China Meteorological Administration, Beijing 100081, China}
\affiliation{Innovation Center for FengYun Meteorological Satellite (FYSIC), Beijing 100081, China}
\email{wangjs@cma.gov.cn}

\author{Xinkai Li}
\affiliation{Changchun Institute of Optics, Fine Mechanics and Physics, Chinese Academy of Sciences, Changchun 130033, China}
\email{lixinkai@ciomp.ac.cn}

\author{Peng Wang}
\affiliation{Changchun Institute of Optics, Fine Mechanics and Physics, Chinese Academy of Sciences, Changchun 130033, China}
\email{sjtuwangpeng@126.com}

\author{Kefei Song}
\affiliation{Changchun Institute of Optics, Fine Mechanics and Physics, Chinese Academy of Sciences, Changchun 130033, China}
\email{songkf@ciomp.ac.cn}

\author{Bo Chen}
\affiliation{Changchun Institute of Optics, Fine Mechanics and Physics, Chinese Academy of Sciences, Changchun 130033, China}
\affiliation{State Key Laboratory of Applied Optics, Changchun 130033, China}
\email{chenb@ciomp.ac.cn}

\correspondingauthor{Chuan Li}
\correspondingauthor{Jinsong Wang}

\begin{abstract}
For flares associated with eruptive filaments, the filament normally rises above the post-flare loops. Here, we report an unusual event in which the post-flare loops instead overlie a filament that remains non-ejected throughout the flare. This event was simultaneously captured by the Chinese H$\alpha$ Solar Explorer (CHASE), the Solar Extreme-UltraViolet Imager (SUVI) onboard Fengyun-4C, the Hard X-ray Imager (HXI) onboard the Advanced Space-based Solar Observatory (ASO-S) and the Solar Dynamics Observatory (SDO). Extreme Ultraviolet (EUV) observations provided by FY4C/SUVI and SDO show that two hot loop systems form at the flare onset. The western end of the northern loop system undergoes a southwestward drifting motion with gradually increasing shear, whereas the southern loop system progressively converges toward and eventually overlies the northern one. Meanwhile, three-dimensional magnetic field extrapolations reveal an increase in both the twist of the filament-supporting magnetic structure and its poloidal magnetic field after the flare. Taken together, these observations suggest that the flare is driven by a self-constraining zipper-type magnetic reconnection, in which highly sheared arcades reconnect to form a series of overlying flare arcades and an underlying flux rope. This reconnection continuously strengthens the magnetic confinement of the filament, preventing its outward eruption, while facilitating substantial plasma drainage along the reconnected magnetic field revealed by the CHASE H$\alpha$ spectroscopic observations.

\end{abstract}


\section{Introduction} 
Solar flares are among the most violent eruptive phenomena in the solar atmosphere. They are characterized by the sudden release of enormous amounts of magnetic energy stored in the corona, producing intense electromagnetic radiation over nearly the entire spectrum, from radio waves to $\gamma$-rays. Typical flares can release energies up to 10$^{32}$ erg within tens of minutes to several hours, accompanied by plasma heating, particle acceleration, and large-scale magnetic restructuring \citep{2011LRSP....8....6S,2017LRSP...14....2B}. As is well known, solar flares are frequently associated with coronal mass ejections (CMEs) and solar energetic particles (SEPs) that may severely disturb the geomagnetic field, causing damage to near-Earth high-tech facilities \citep{2007JGRA..11210102Z,2009RaSc...44.0A17T}. Thus, investigating the physical mechanism and evolution of solar flares is crucial for both solar physics and space weather forecasting.

Solar filaments, observed as prominences above the solar limb, are cool and dense plasma structures suspended in corona. They are typically located above polarity inversion lines and are believed to be supported by highly sheared magnetic arcades or magnetic flux ropes \citep{2014LRSP...11....1P}. Observational and theoretical studies suggest that the relationship between flares and filaments is intrinsically coupled \citep{2000JGR...105.2375L,2004ApJ...614.1054J,2011MNRAS.414.2803Y,2011LRSP....8....1C}. Based on the standard flare model (commonly referred to as the CSHKP model; \citealt{1964NASSP..50..451C,1968IAUS...35..471S,1974SoPh...34..323H,1976SoPh...50...85K}), a filament or a flux rope structure erupts due to loss of equilibrium, stretching the overlying field lines and forming a current sheet beneath it, where magnetic reconnection produces plasma heating in the lower atmosphere and further propels the erupting structure outward. This model provides a coherent framework for understanding the physical connection between filament eruptions, solar flares and CMEs.

However, not all filament eruptions can successfully escape from the Sun and evolve into CMEs \citep{2004ApJ...614.1054J,2005A&A...435.1149V,2011MNRAS.414.2803Y}. Depending on the fraction of the escaped filament material, filament eruptions can be classified as full, partial, or failed eruptions \citep{2007SoPh..245..287G}. The full and partial eruptions are generally associated with CMEs, whereas the failed ones finally fall back to the solar surface after propagating to a certain height without any counterparts observed in the white-light coronagraph. The failure of filament eruptions can be attributed to two main classes of physical mechanisms from the perspective of force analysis. One is the insufficiency or weakening of the upward driving force, which may result from torus-unstable but weakly twisted flux ropes~\citep{2015Natur.528..526M}, the loss of poloidal flux via reconnection above the flux rope~\citep{2023ApJ...951L..35C,2025ApJ...979..113L,2026NatAs.tmp..107G}, or the redirection of the non-axisymmetry induced force~\citep{2021NatCo..12.2734Z}. The other is strong or enhanced downward Lorentz force exerted by the external magnetic field, such as the best-known strapping force~\citep{2006PhRvL..96y5002K,2007ApJ...665.1428W,2010ApJ...725L..38G,2017ApJ...843L...9W,2025ApJ...991..184Q}, and the toroidal magnetic cage induced force~\citep{2007ApJ...668.1232F,2026A&A...710A.212G}.

The physical mechanisms responsible for filament eruptions can be broadly classified into two categories. The first comprises reconnection-driven mechanisms, including the tether-cutting model~\citep{2001ApJ...552..833M}, the breakout model~\citep{1999ApJ...510..485A}, the emerging-flux trigger model~\citep{2000ApJ...545..524C}, the zipper reconnection model~\citep{2017SoPh..292...25P} and so on. The second involves magnetohydrodynamic (MHD) instabilities, primarily the torus instability~\citep{2006PhRvL..96y5002K} and the kink instability~\citep{2005ApJ...630L..97T}. Despite their different triggering mechanisms, these models share a common evolutionary picture consistent with the standard CSHKP flare model: once the filament or its supporting flux rope loses equilibrium, it erupts outward, stretching the overlying magnetic field and inducing magnetic reconnection beneath it, which produces post-flare arcades below the erupting filament.

However, we identified an intriguing flare on 2026 March 1, in which post-flare loops formed above a filament that shows no features of ejection throughout the entire eruption. This unusual configuration is inconsistent with the standard eruptive flare scenario and raises new questions about the magnetic coupling between flare reconnection and filament stability. In this Letter, we combine multiwavelength EUV observations, H$\alpha$ spectroscopic measurements of the filament plasma, and nonlinear force-free field extrapolations to investigate the magnetic evolution of this event and explore the physical mechanism responsible for the non-ejected filament. The data and methods employed in this letter are briefly described in Section~\ref{sec:data_method}. The results are presented in Section~\ref{sec:res}, followed by the discussion and conclusion in Section~\ref{sec:conclu}.

\section{Observations and Methods}\label{sec:data_method}
The flare of interest occurred on 2026 March 1 in NOAA active region (AR) 14380. This active region was located at approximately S20° E50° and exhibited a classic four-polarity configuration (Figure~\ref{fig:suvi}). The flare took place at the central PIL between N1 and P2 of AR 14380, which is depicted by the orange curve in Figure~\ref{fig:suvi}(b), and had a magnitude of C6.1. According to the GOES 1–8 Å soft X-ray (SXR) light curve (Figure~\ref{fig:suvi}(c)), this flare started at 01:55 UT, peaked at 02:06 UT, and ended at 02:22 UT.

The flare was well observed by the Chinese H$\alpha$ Solar Explorer (CHASE; \citealt{2022SCPMA..6589602L}) over the time interval from 01:54 to 02:22 UT, covering the entire eruption process. The CHASE mission provides full-disk high spectral-resolution spectra in the H$\alpha$ and Fe I wavebands \citep{2022SCPMA..6589603Q}, which can be used to derive the chromospheric and photospheric dopplergrams, respectively. The CHASE data during the flare were acquired in the binning mode, achieving a spectral resolution of 0.0484~\AA~per pixel and a spatial sampling of 1.04\arcsec~per pixel (twice the normal resolution), with a cadence of 71 s.

In addition, we adopted EUV observations from the Solar Extreme-UltraViolet Imager (SUVI) onboard FY-4C to investigate the brightening evolution. FY-4C is the third satellite in China's second‑generation geostationary meteorological satellite series. SUVI provides observations in four EUV passbands, namely 94 Å (Fe XVIII), 171 Å (Fe IX), 211 Å (Fe XIV), and 304 Å (He II). For each passband, images are acquired with three exposure settings (long, medium, and short) to accommodate a wide range of coronal and chromospheric emission intensities. The instrument has a spatial sampling of approximately 1.24\arcsec~per pixel and a conventional cadence of about 1 minute. When operated in a single-passband observing mode, the cadence can be increased to as high as 5 s.

We also employed EUV observations from the Atmospheric Imaging Assembly (AIA; \citealt{2012SoPh..275...17L}) to perform the Differential Emission Measure (DEM) analysis, and photospheric vector magnetograms from the Helioseismic and Magnetic Imager (HMI; \citealt{2012SoPh..275..229S}) for coronal magnetic field extrapolation. Both instruments are onboard the Solar Dynamics Observatory (SDO; \citealt{2012SoPh..275....3P}). AIA provides full-disk EUV images with a spatial sampling of 0.6\arcsec~pixel$^{-1}$ and a cadence of 12 s, whereas HMI provides full-disk vector magnetic field measurements with a spatial sampling of 0.5\arcsec~pixel$^{-1}$ and a cadence of 720 s. The hard X-ray (HXR) sources were derived from the data of the Hard X-ray Imager (HXI; \citealt{2019RAA....19..160Z,2024SoPh..299..153S}) aboard the Advanced Space-based Solar Observatory (ASO-S; \citealt{2019RAA....19..156G}). HXI provides HXR images and spectra from the Earth’s perspective in the energy range from 10 to 300 keV, with a spatial resolution as high as 3.1\arcsec~pixel$^{-1}$. Besides, the time resolution is 4 seconds in normal mode and can be as high as 0.125 seconds in burst mode. 

\setcounter{footnote}{0}
To understand the distribution of plasma at different temperatures during the flare, we reconstructed DEM using the regularized inversion method developed by Hannah \& Kontar \citep{2012A&A...539A.146H,2013A&A...553A..10H}\footnote{The DEM code is publicly available at: https://github.com/ianan/demreg}. This method employs Generalized Singular Value Decomposition (GSVD) to efficiently solve the regularized inversion problem and recover the plasma emission measure as a function of temperature. The DEM reconstruction was performed using simultaneous observations from the six optically thin EUV channels of SDO/AIA (94, 131, 171, 193, 211, and 335~\AA), which together provide sensitivity to plasma over a broad temperature range from approximately 0.5 to 20 MK.

We estimated the Doppler shift by the moment method, which can be expressed by $v_{\rm{D}} = \frac{\sum v_{\rm{i}} (I_{\rm{i}} - I_{\rm{c}})}{\sum (I_{\rm{i}} - I_{\rm{c}})}$, where $v_{\rm{D}}$ denotes the Doppler velocity, $I_{\rm{c}}$ is the continuum intensity, and the quantities with subscript ``i" represent the corresponding Doppler velocity and intensity at i-th wavelength point, respectively. The reference line-core wavelength, which refers to zero Doppler shift, is usually obtained from the averaged spectral profile of a nearby quiescent region to correct for systematic wavelength offsets caused by solar rotation, spacecraft motion, and instrumental effects.

The three-dimensional non-linear force-free field (NLFFF) extrapolation was implemented by the magneto-frictional module \citep{2016ApJ...828...82G} in the Message Passing Interface Adaptive Mesh Refinement Versatile Advection Code (MPI-AMRVAC; \citealt{2018ApJS..234...30X,2023A&A...673A..66K}). The magneto-frictional method iteratively solves the magnetic induction equation, enabling the initial potential field to progressively converge to an NLFFF that best matches the vector magnetic field prescribed at the bottom boundary. As the lower boundary condition, we adopted the HMI full-disk vector magnetograms obtained at 01:36 UT and 02:36 UT, respectively. Before inputting into the calculation, the vector magnetic field data were processed with the 180$^\circ$ ambiguity elimination of the transverse component and subsequently transformed into a local Cartesian coordinate system using the rigid-rotation method \citep{1990SoPh..126...21G}.

\section{Results}\label{sec:res}
\subsection{General Evolution}
AR 14380 exhibited a typical quadrupolar magnetic configuration and was located in the southeast of the solar disk on 2026 March 1. The flare we investigated occurred near the central PIL of AR 14380. We checked the quick-look images from LASCO aboard SOHO and found no CME counterpart in white-light coronagraph observations. During the flare, the peak of the GOES SXR flux derivative (the gray dashed line in Figure~\ref{fig:suvi}(c)) closely matches the peak of the background-subtracted 12–23 keV HXR light curve (the red solid line in Figure~\ref{fig:suvi}(c)) observed by ASO-S/HXI, consistent with the Neupert effect~\citep{1968ApJ...153L..59N}. This correspondence suggests that the flare heating was primarily driven by non-thermal electrons.

In the FY-4C/SUVI 94~\AA\ observations (Figures~\ref{fig:suvi}(d--f)), two hot loop systems are formed on the northern and southern sides of the flaring region at the onset of the flare. During the subsequent evolution, the western end of the northern loop system drifts southwestward, causing the loops to become progressively more sheared, while the southern loop system migrates northeastward. Eventually, the two loop systems overlap in projection, with the southern loop system overlying the northern one. We made a slice intersecting both loop systems toward the southwest (the orange dashed line in Figure~\ref{fig:suvi}(e)) and constructed a time--distance {diagram (see Figure~\ref{fig:suvi}(i)). It reveals that the drift velocities of the northern and southern loop systems are approximately 23.9 and $-21.5$ km s$^{-1}$, respectively. The opposite motions indicate a pronounced converging evolution of these two loop systems.

The filament associated with the flare was located near the center of the active region and approximately aligned with the central PIL, hereinafter referred to as the central filament. Revealed by the EUV observations (Figure~\ref{fig:aia}), the central filament remained confined throughout the flare, showing no signs of rising or material ejection in the plane of the sky, while the flare loops were consistently formed above it. To assess its evolution, we compared the post-flare filament at 02:33 UT with its pre‑flare morphology at 01:48 UT (see the filament structure and white dotted line in Figure~\ref{fig:suvi}(h)). The comparison reveals no significant changes in either its overall morphology or its position, indicating that the filament remained non-ejected despite the intense flare activity.

Although the central filament remained non-ejected all the time, the small filament east of it eventually lost equilibrium and erupted under the strong perturbation of the flare, as revealed by Figure~\ref{fig:aia}. The eastern small filament began to eject eastward at 02:02:21 UT, approximately 7 minutes after the flare onset (marked by the vertical green dashed line in Figure~\ref{fig:suvi}(c)). This filament ejection is most evident in the 171 Å running-difference images, but is barely discernible in high-temperature channels (e.g., 94 and 131 Å). This behavior is consistent with the million--Kelvin warm channel ejection reported by \cite{2024ApJ...967..130L}.

Figures~\ref{fig:dem}(a-d) present the distribution of the DEM-weighted temperatures, which is defined as $T_{\text{DEM}} = \frac{\sum\limits_{j} T_j \text{DEM}_j}{\sum\limits_{j} \text{DEM}_j}$ where $\text{T}_j$ is the center temperature of the $j$-th temperature bin, and $\text{DEM}_j$ is the corresponding differential emission measure \citep{2002ApJ...578L.161S}. The DEM-weighted temperature maps also reveal the pronounced southward drift of the western end of the northern loop systems, which gradually increases the shear of these loops. During the flare, the coronal plasma above the central filament was heated from 1 MK to about 10 MK (Figure~\ref{fig:dem}(f)), whereas the outer boundary of the filament channel was heated to temperatures of approximately 1.0–1.5 MK (Figure~\ref{fig:dem}(e)).

\subsection{Spectral Analysis }
CHASE/HIS captured the entire evolution of the flare, allowing us to investigate the Doppler motions associated with both the flare and the co-spatial filament using H$\alpha$ spectroscopic observations. The H$\alpha$ line-core images show that the filament exhibits no significant displacement in the plane of the sky throughout the event, and no mass ejection is detected (Figures~\ref{fig:chase}(a--c)). The plasma in the lower atmosphere on both sides of the filament was heated, producing two nearly parallel flare ribbons, with the northwestern one extending along the filament and forming a hook--shaped brightening. During the early phase of the flare, the flare ribbons exhibit redshifts of approximately 10 km s$^{-1}$ (Figure~\ref{fig:chase}(e)), consistent with the signatures of chromospheric condensation. The redshifted flare ribbons are spatially coincident with the ASO-S/HXI 12–23 keV source (Figure~\ref{fig:chase}(b)), indicating that non-thermal electrons played an important role in heating the lower atmospheric plasma and causing the associated condensation.

Although the filament exhibits little apparent motion in imaging observations, its internal plasma undergoes significant dynamical evolution. As shown in Figure~\ref{fig:chase}(d), at the onset of the flare, the two filament legs exhibit slight redshifted flows, whereas the region near the filament apex is characterized by blueshifted flows. As the flare progresses, the internal plasma motions intensify, with Doppler velocities reaching approximately 30 km s$^{-1}$ during the middle and late phases of the flare (Figure~\ref{fig:chase}(f)). This Doppler pattern indicates substantial mass transport and drainage within the filament channel, likely driven by magnetic reconnection involving the filament-supporting magnetic field.

Because the filament appears as an absorption feature in the H$\alpha$ spectra, we normalized the spectra to the continuum intensity and calculated the average normalized intensity within $\pm$0.5~\AA\ of the reference line-core wavelength determined from the background region. By applying an appropriate intensity threshold, the filament morphology was identified in the CHASE H$\alpha$ spectral observations obtained before and after the flare (Figures~\ref{fig:chase}(g-h)). The post-flare filament occupies a noticeably smaller area than its pre-flare counterpart, although its overall morphology remains roughly unchanged, indicating a reduction in the amount of absorbing filament material.

We further calculated the average normalized H$\alpha$ line profiles over the filament region before and after the flare. Compared with the reference profile derived from the surrounding background region, the filament profiles are significantly broadened but exhibit no obvious net Doppler shift, likely owing to the superposition of unresolved internal mass motions along the line of sight. The gray dashed curve in Figure~\ref{fig:chase}(i) shows the difference between the post-flare and pre-flare profiles, magnified by a factor of 60 for clarity. The post-flare profile exhibits weaker absorption than the pre-flare profile, implying a decrease in the optical thickness and column density of the filament after the C6.1 flare. This result is consistent with the observed mass drainage inferred from the Doppler measurements.

\subsection{Magnetic Field}
We extrapolated the three-dimensional coronal magnetic field using the NLFFF model based on the vector photospheric magnetograms obtained at 01:36 UT and 02:36 UT, corresponding to the pre-flare and post-flare stages, respectively. Magnetic field lines traced along the central PIL beneath the filament reveal a bundle of strongly sheared magnetic arcades before the flare (Figure~\ref{fig:mag}(a)), whereas the post-flare extrapolation exhibits a coherent flux-rope configuration (Figure~\ref{fig:mag}(c)). This transition suggests that the magnetic reconnection related to this flare might transform the sheared arcade into a more twisted magnetic structure supporting the central filament. 

In addition, we calculated the poloidal component of the potential field above the central filament. Because the central filament exhibited no significant upward motion, the poloidal direction was defined as the direction perpendicular to both the axis of the magnetic structures supporting the filament and the vertical direction, satisfying $\mathbf{e}_{\rm p} \times (\mathbf{e}_{\rm axis} \times \mathbf{e}_{\rm z}) = 0$. We then constructed a slice approximately perpendicular to the axis of the sheared arcades and the flux rope (white straight lines in Figures~\ref{fig:mag}(a) and (c)) and calculated the poloidal magnetic field within this plane over heights ranging from $70''$ to $170''$ (corresponding to 50--122 Mm). Figure~\ref{fig:mag}(b) compares the distributions of the poloidal field at 01:36 UT and 02:36 UT. The contours corresponding to the same field strength are displaced upward after the flare, indicating an overall enhancement of the poloidal field above the filament.  

\section{Discussion and Conclusion}\label{sec:conclu}
In this study, we investigate an event in which a filament, co-spatial with a flare, remained confined throughout the entire flare evolution, while post-flare loops progressively formed above it. A similar configuration has previously been reported only in partially erupted double-decker filaments, where the upper branch erupts and the magnetic reconnection underneath produces post-flare loops above the surviving lower branch~\citep{2019ApJ...875...71Z,2021ApJ...923..142C,2023ApJ...959...69H}. However, for the event investigated here, although the C6.1 flare destabilized the nearby eastern small filament, leading to a warm-channel eruption~\citep{2024ApJ...967..130L} 7 minutes after the flare onset, the central filament spatially coincident with the flare withstood the strong disturbance and exhibited no evident mass ejection throughout the event in either the EUV or H$\alpha$ observations. During the flare, the post-flare loops constantly formed above the central confined filament, challenging the conventional picture that post-flare arcades are produced beneath an erupting filament or flux rope. 

The EUV observations and DEM-weighted temperature maps reveal two distinct hot loop systems associated with the flare. During the flare evolution, the western end of the northern loop system gradually drifts southwestward, with an apparent velocity of approximately 23.9 km s$^{-1}$. Meanwhile, the southern loop system migrates northeastward at a velocity of approximately $-21.5$ km s$^{-1}$ and is located at a higher altitude than the northern loop system. In the late phase of the flare, the two loop systems eventually overlap in projection. A comparison of the extrapolated three-dimensional magnetic fields before and after the flare reveals that the filament-supporting magnetic structure evolves from a set of highly sheared arcades into a coherent flux rope-like configuration, implying an increase in magnetic twist. Furthermore, the poloidal component of the magnetic field is strengthened after the flare, probably due to the progressive formation of newly reconnected loops above the filament. These results suggest that the flare reconnection not only increases the twist of the filament-supporting magnetic field but also enhances the magnetic confinement exerted by the overlying field.

It is common for the twist of the filament-supporting magnetic structure to increase during filament eruptions~\citep{2017ApJ...839..128W,2017NatCo...8.1330W,2023ApJ...944..161S}. One possible explanation for this behavior is the zipper reconnection model~\citep{2017SoPh..292...25P}, in which the reconnection front propagates progressively along the PIL, converting highly sheared arcades into an overlying twisted flux rope and underlying flare arcades. This process gradually builds up twist in the erupting flux rope and sets the stage for the subsequent main‑phase reconnection that expands perpendicular to the PIL. The drifting motion of the two hot loop systems resembles the progressive propagation expected in zipper reconnection. However, unlike the standard zipper-reconnection scenario, the filament remained non-ejected throughout the entire flare. Moreover, the newly formed, more twisted field lines are located beneath the post-flare loops rather than erupting outward. Consequently, the filament-supporting magnetic structure becomes progressively more twisted, while the overlying magnetic confinement is simultaneously strengthened by the newly reconnected arcades. These observations suggest that the C6.1 flare on 2026 March 1 might be governed by a self-constrained zipper-type magnetic reconnection process (Figure~\ref{fig:mag}(d)), in which flare reconnection not only restructures the filament-supporting magnetic field but also continuously reinforces the overlying magnetic cage, thereby preventing the supporting structure and the filament from erupting.

Magnetic reconnection involving the filament-supporting field can effectively heat the filament plasma to higher temperatures~\citep{2012ApJ...761...62C,2025A&A...698A.301J}. In the present event, the DEM analysis reveals that the outer boundary of the filament was heated to approximately 1.0--1.5 MK, which perhaps results from the self-constrained zipper-type reconnection. In addition, this reconnection reconfigured the filament-supporting magnetic structure and strongly perturbed the filament plasma, driving substantial mass transport and drainage within the filament channel. Part of the filament material drained toward the solar surface under gravity, producing redshift in the two legs and blueshift in between. As the flare progressed, the Doppler velocities of this flow increased from approximately 10 km s$^{-1}$ to about 30 km s$^{-1}$. Although the overall morphology of the filament was nearly unchanged, both the reduction in the projected filament area and the increase in the mean H$\alpha$ intensity indicate a significant decline of absorbing material within the filament channel.

In summary, we explored the C6.1 flare that occurred in NOAA AR 14380 on 2026 March 1. By combining EUV observations with three-dimensional magnetic field extrapolations, we propose that this event is dominated by a self-constrained zipper-type magnetic reconnection process, in which highly sheared field lines reconnect progressively to form an underlying flux rope and overlying post-flare arcades. As a result, the twist of the filament-supporting magnetic structure increases during the flare. But this supporting structure and filament remain non-ejected throughout the event because the confinement exerted by the overlying magnetic field is simultaneously reinforced. Meanwhile, the reconfiguration of the filament-supporting magnetic field drives substantial mass drainage within the filament channel. Our results therefore provide new insight into the coupling between flare reconnection, filament eruption, and magnetic confinement, highlighting a previously unrecognized pathway by which flare reconnection can simultaneously enhance the twist and confinement of the filament‑supporting magnetic system.

\begin{acknowledgments}
We sincerely appreciate the anonymous reviewer for the valuable suggestions and comments. We are grateful to the CHASE, ASO-S/HXI, SDO/AIA, and HMI science teams, as well as the Post Launch Test team of the CMA FY-4 Program, for providing the calibrated data used in this paper. CHASE is a space mission supported by the China National Space Administration (CNSA). The FY-4C satellite is operated by the National Satellite Meteorological Center (NSMC). The ASO-S mission is supported by the Strategic Priority Research Program on Space Science of the Chinese Academy of Sciences (CAS). SDO is a mission of NASA's Living With a Star Program. The authors are funded by Basic Research Program of Jiangsu (BK20251209), NSFC (12333009) and the Fundamental Research Funds for the Central Universities (KG202506).

\end{acknowledgments}

\bibliography{20260301}
\bibliographystyle{aasjournalv7}

\begin{figure}[ht!]
	\centering
	\includegraphics[width=1\textwidth]{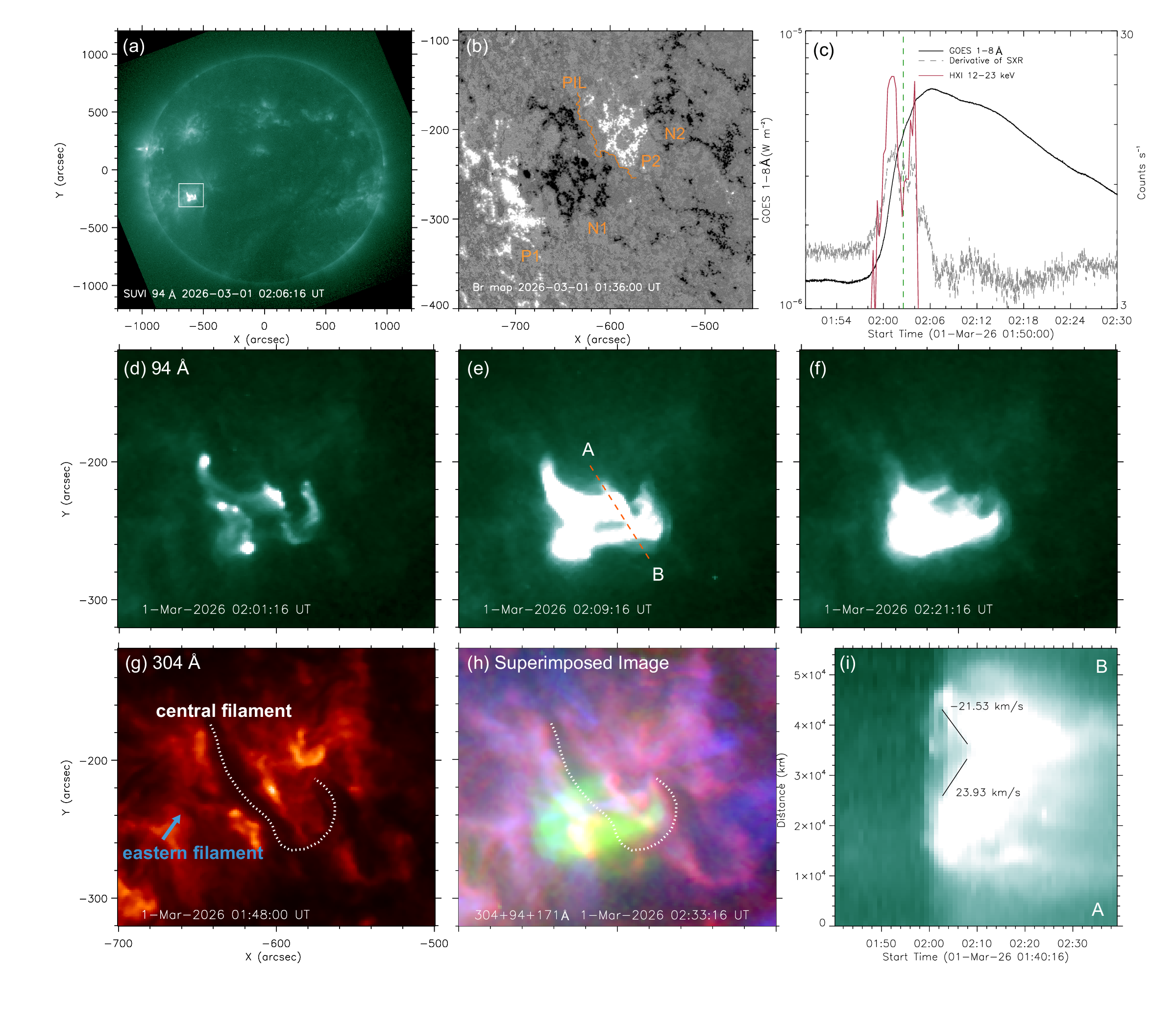}
	\caption{Overview~and EUV evolution of the C6.1 flare on 2026 March 1. (a) FY-4C/SUVI 94~\AA~full-disk image at 02:06:16 UT. The white rectangle marks the region of interest encompassing the flare and AR 14380. (b) Photospheric radial magnetic field map inferred from the SDO/HMI vector magnetogram prior to flare onset at 01:36 UT. The orange curve denotes the central PIL of the active region. (c) Soft X-ray (SXR) and hard X-ray (HXR) light curves during the C6.1 flare. The black solid line represents the SXR flux, with the gray dashed line showing its time derivative. The red solid line indicates the evolution of the 12–23 keV HXR count rate observed by ASO-S/HXI. The green vertical dashed line marks the onset of the ejection of the small filament east of the central filament. (d--f) FY-4C/SUVI 94~\AA~images at three times. The orange dashed line marks the slice for the time-distance diagram. (g) SUVI 304~\AA~image at 01:48 UT, before the onset of the flare. (h) Composite image of the three SUVI EUV channels at 02:33 UT. The white dotted lines in (g) and (h) indicate the filament spine measured from the 304~\AA~image at 01:48 UT. (i) Time-distance diagram acquired from the SUVI 94~\AA~observations. A 5--second online animation depicting the evolution of the flare in SUVI EUV observations is presented.} \label{fig:suvi}
\end{figure}

\begin{figure}[ht!]
	\centering
	\includegraphics[width=1\textwidth]{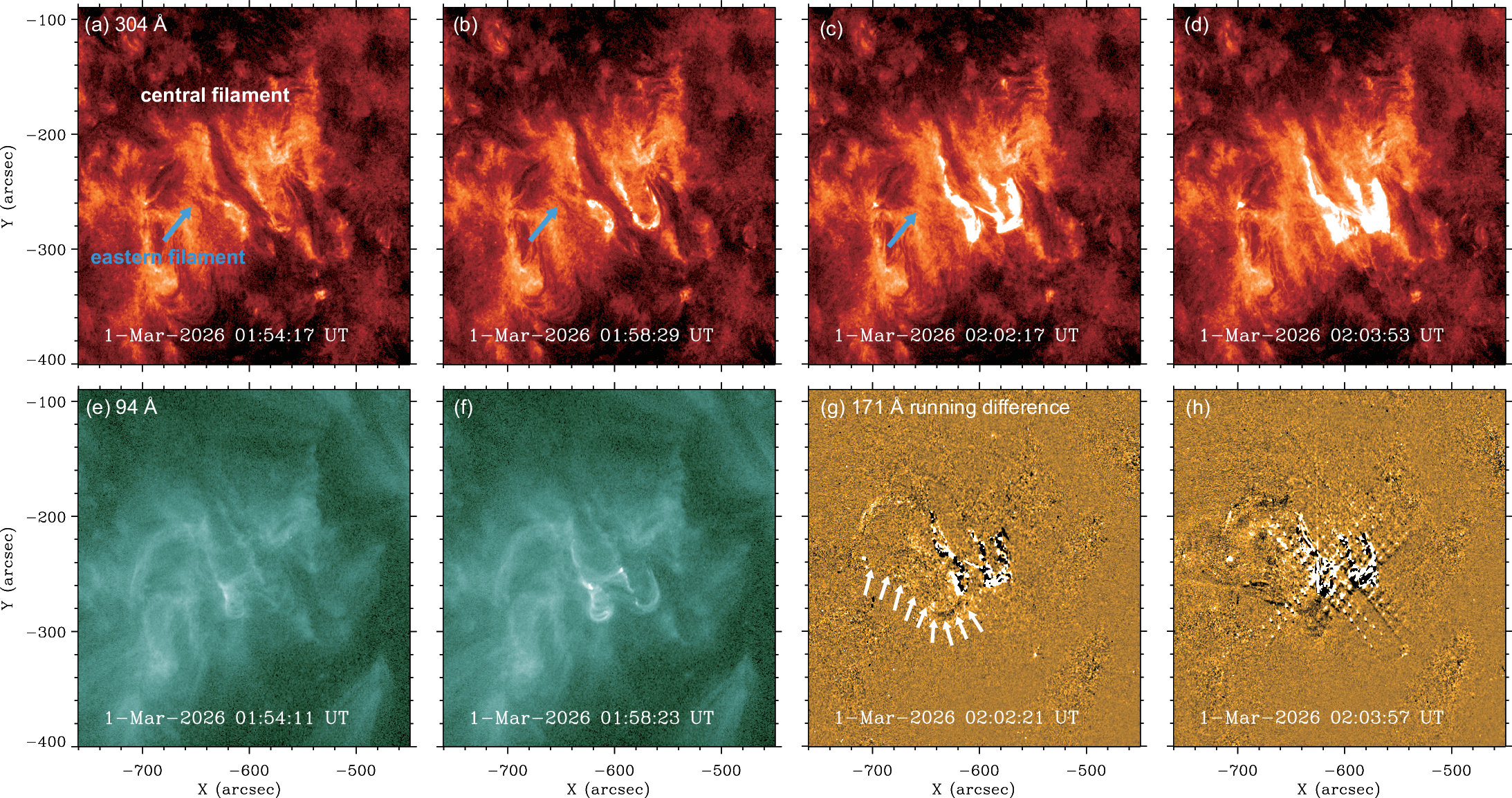}
	\caption{SDO/AIA EUV observations of the non-ejected central filament, and the ejection of the eastern small filament disturbed by the nearby C6.1 flare. (a--d) 304~\AA~images at four selected times. The blue arrows point to the eastern small filament. (e--f) 94~\AA~images before the flare and at its onset, respectively. (g--h) 171~\AA~running--difference images showing the ejection of the small filament. The white arrows in (g) outline the leading-edge structure of the filament ejection. A 19 s animation is available online, showing the evolution of the flare and the ejection of the small filament in SDO/AIA 304 Å images and 171 Å running-difference images.} \label{fig:aia} 
\end{figure}

\begin{figure}[ht!]
	\centering
	\includegraphics[width=1\textwidth]{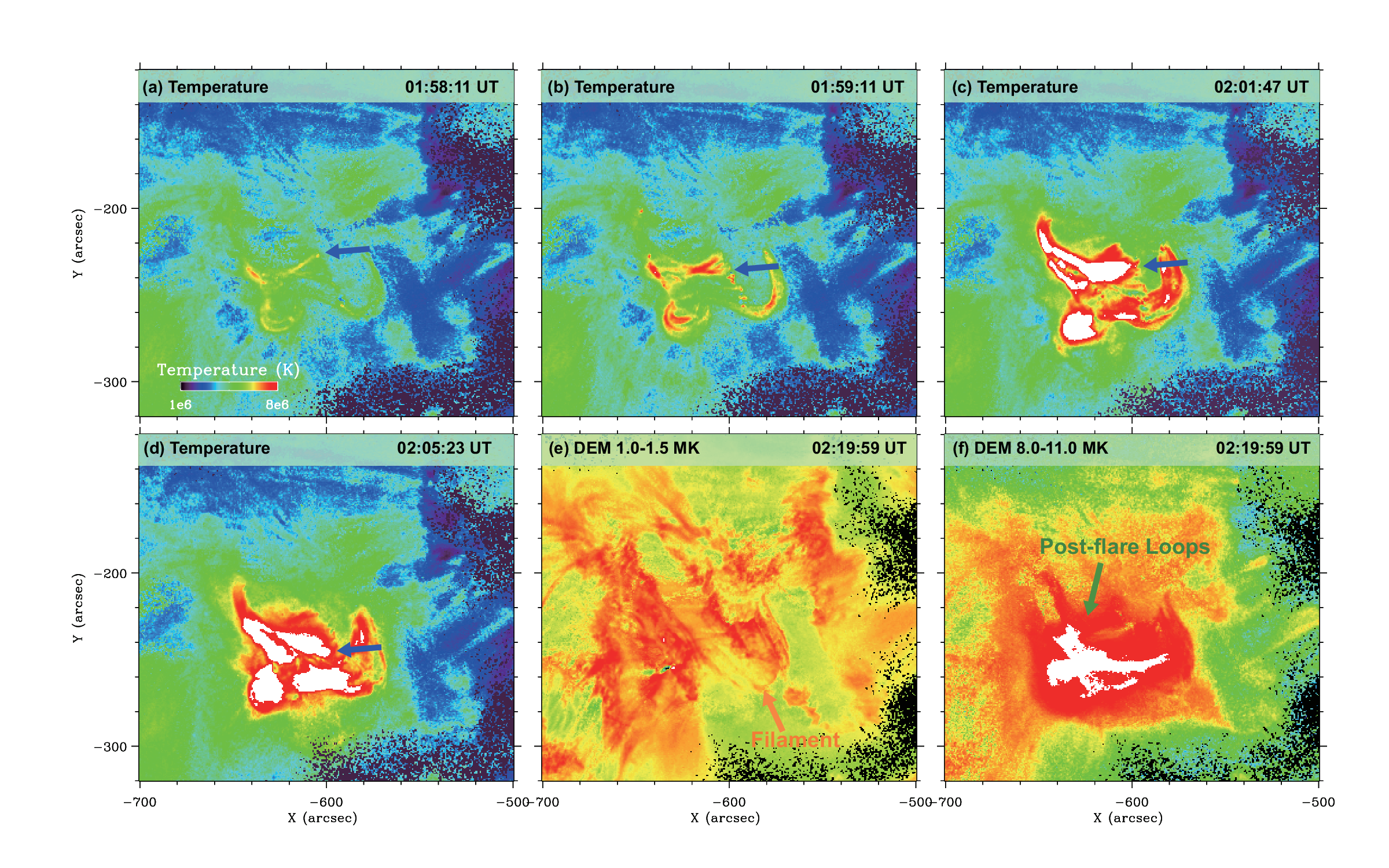}
	\caption{Computed DEM results. (a--d) Temporal evolution of the DEM--weighted temperature distribution. The dark blue arrows point to the drifting end of the northern heated loops. (e--f) DEM images at 02:19:59 UT showing the coronal plasma in the temperature ranges of 1.0--1.5 MK and 8.0--11.0 MK, respectively. } \label{fig:dem}
\end{figure}

\begin{figure}[ht!]
	\centering
	\includegraphics[width=1\textwidth]{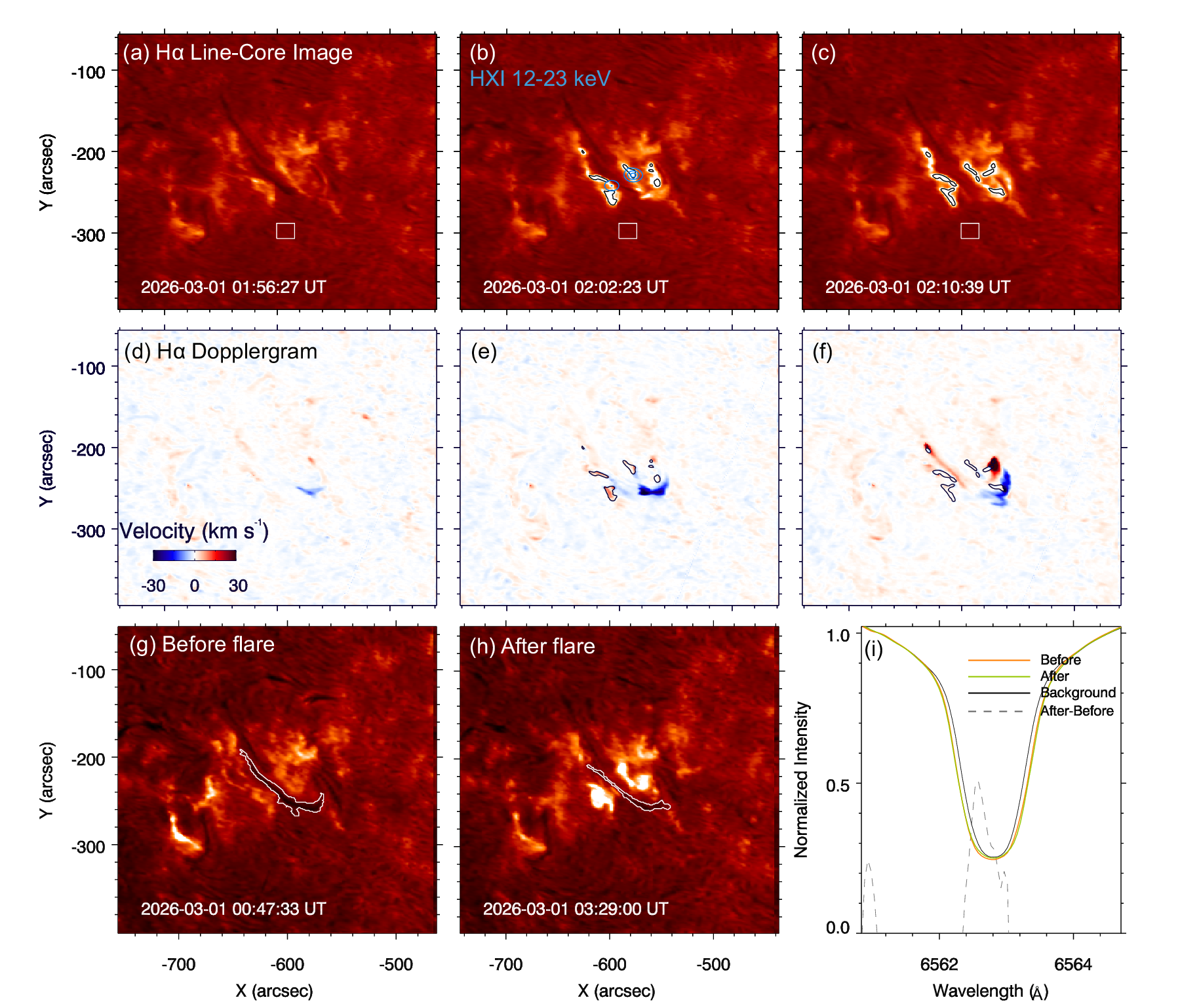}
	\caption{CHASE/HIS observations of the flare. (a--c) H$\alpha$ line--core images showing the evolution of the flare. The black contours indicate the locations of the flare ribbons. The blue curves denote the 10\%, 40\%, and 80\% levels of the maximum of the reconstructed HXR sources in the 12--23 keV energy band. The white rectangles mark the background regions used to define the reference zero-shift spectral profile. (d--f) Chromospheric dopplergrams of the flare, with positive values representing redshifts and negative values representing blueshifts. The black contours outline the same flare ribbons as in panels (b--c). (g--h) CHASE/HIS H$\alpha$ line--core images just before and after the flare. The white solid lines trace the filament identified from the spectral line. (i) Comparison of the averaged spectral line profiles of the filament before (orange line) and after (green line) the flare. The black solid line represents the averaged line profile of the background region. The gray dashed line shows the difference between the line profiles after and before the flare. A 2--second animation is available online, which includes the H$\alpha$ line-core images and Dopplergrams.} \label{fig:chase}
\end{figure}

\begin{figure}[ht!]
	\centering
	\includegraphics[width=1\textwidth]{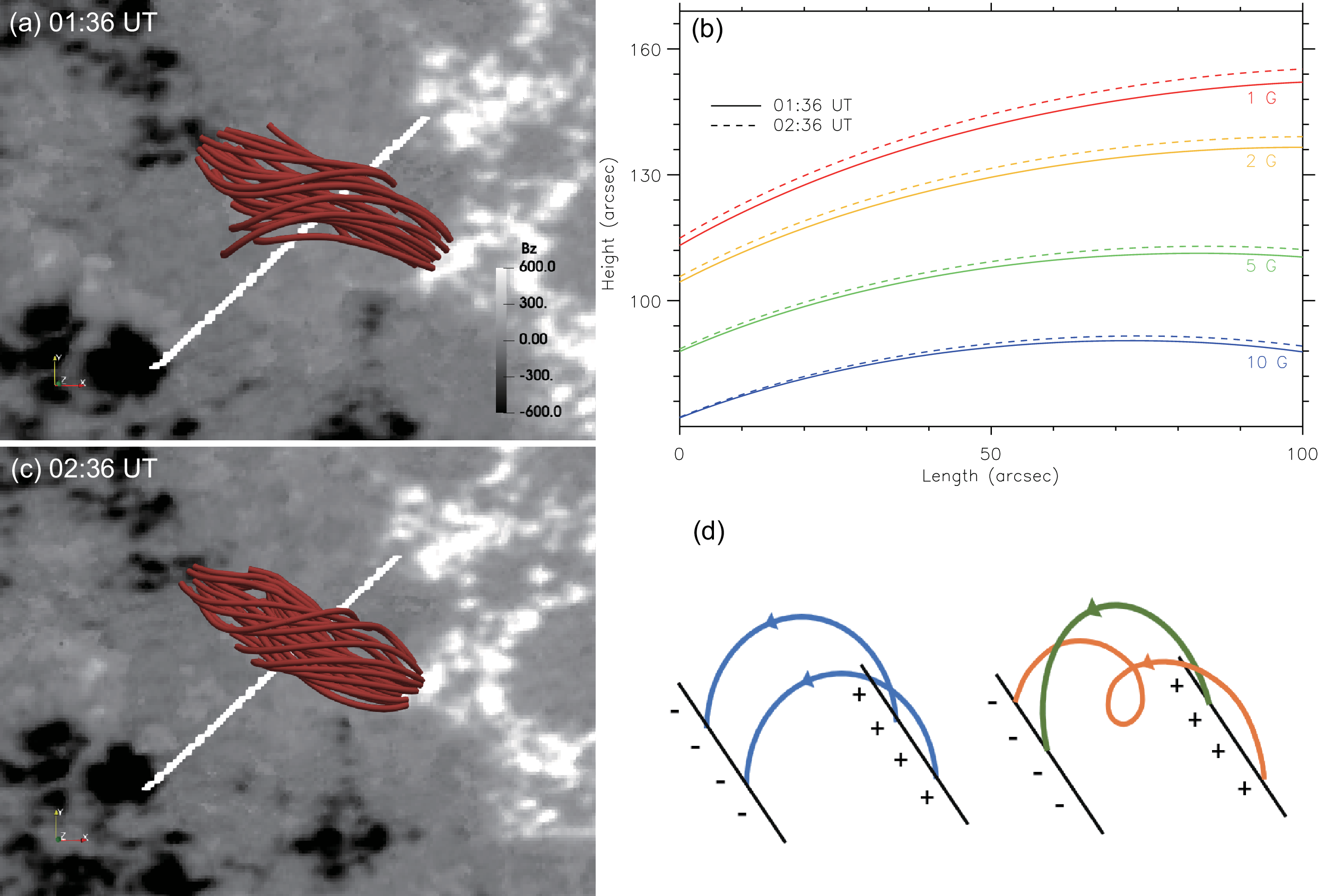}
	\caption{Three--dimensional coronal magnetic field of the active region. Panels (a) and (c) show the non--potential structures around the central PIL at 01:36 and 02:36 UT, overlaid with the B$_{\rm z}$ map. The white straight lines mark the locations used to calculate the poloidal magnetic field above the non--potential structures. (b) Comparison of the poloidal magnetic field distributions before and after the flare. The colored solid lines denote the contours of the poloidal magnetic field at 01:36 UT, whereas the colored dashed lines show the distributions at 02:36 UT. (d) Schematic diagram of the magnetic reconnection process in this event. The blue arrowed lines indicate the sheared magnetic arcades before magnetic reconnection. The green and orange arrowed lines represent the flare loop and twisted magnetic field line formed after the reconnection, respectively. } \label{fig:mag}
\end{figure}

\end{document}